\documentclass{moriond}

\usepackage{bm}
\def\planck{\textit{Planck}}
\def\Mcmblens{M_{\rm CMBlens}}
\def\Msz{M_{\rm SZ}}
\providecommand{\Omm}{\Omega_{\mathrm{m}}}

\def\be{\begin{equation}}
\def\ee{\end{equation}}
\def\bea{\begin{eqnarray}}
\def\eea{\end{eqnarray}}

\newcommand{\Photo}{\includegraphics[height=35mm]{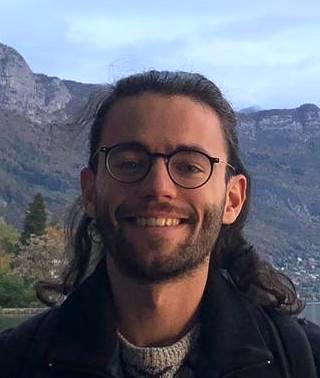}}

\begin{document}
\vspace*{4cm}
\title{Measuring cluster masses with Planck and SPT-SZ CMB lensing}

\author{ Alexandre Huchet }

\address{Université Paris-Saclay, CEA, IRFU, 91191, Gif-sur-Yvette, France}

\maketitle\abstracts{Galaxy cluster masses help to constrain cosmological parameters through the halo mass function. To get rid of major biases in the mass measurement, we directly probe the cluster gravitational potentials by observing their gravitational lensing on the Cosmic Microwave Background (CMB). We measured the average mass of a 468-cluster sample using SPT-SZ and \planck{}/HFI PR2 sky maps and compared it with the masses provided in the SPT-SZ data set. We found an average mass ratio of $\langle \Mcmblens/\Msz \rangle = 0.98 \pm 0.19 \rm{~(stat.)} \pm 0.03 \rm{~(syst.)}$~\cite{huchet_measurement_2024}, in agreement with SPT-SZ masses. We also showed that the CMB large scales are important for the reconstruction of the lensing potential of a cluster with a quadratic estimator.
}

\section{Using CMB lensing to measure cluster masses}

Galaxy clusters are the largest gravitationally bound structures. Their distribution in mass and redshift is very sensitive to cosmological parameters, particularly $\Omm$ and $\sigma_8$, and can thus be used to constrain them. Redshifts are obtained with optical or spectroscopic surveys. Masses measurements can be obtained by various methods and are subject to biases. Galaxy clusters actually contain only $\sim 3\%$ of galaxies, $\sim 12\%$ of hot intracluster gas and $\sim 85\%$ of dark matter. An optical or X-ray survey can map with good accuracy the galaxies or the hot gas, but it will lack most of the mass, which is dark matter that does not emit any light. To compensate this issue, one can use scaling relations that link the mass measured through optical or X-rays to the total mass based on hydro-dynamical simulations. One problem is that, to obtain a mass, some assumptions need to be made, such as having a system in hydro-static equilibrium -- gravitation and pressure perfectly compensating each other. These assumptions are idealistic and create biases in the final measurement.

To avoid those biases, one can choose to probe directly the gravitational potentials of the clusters. This way, $100\%$ of the cluster mass can be obtained by observing how the cluster distorts background images through gravitational lensing. Two different types of background sources can be used: the most used ones are galaxies, but one needs many galaxies to be able to reconstruct the gravitational potential from statistical distortions of galaxy images; the other one is the CMB. We chose the latter and focused on the gravitational lensing of the CMB by galaxy clusters. Indeed, it gets harder to find background galaxies at a redshift $z>1$, whereas the CMB was emitted at a redshift $z \sim 1100$ and will always be behind the observed clusters. Moreover, the signal-to-noise ratio (SNR) is expected to be constant in the redshift range $0.5<z<2$~\cite{melin_measuring_2015}. The gravitational lensing correlates Fourier modes that were independent in the primary CMB. This is how we can manage to retrieve the gravitational potential from this observable.

We used two complementary data sets in this analysis: SPT-SZ, a ground-based telescope and \planck{}, a space-based telescope. On the one hand, the atmosphere of the Earth does not allow the observation of the large scales of the CMB because it fluctuates much faster than the telescope scanning speed. \planck\ can provide these large scales, whereas the SPT-SZ data suffers from an anisotropic transfer function that describes which CMB scales are lost. On the other hand, it is hard to send large mirrors in space that would allow for the observation of the CMB small scales. SPT-SZ and its 10-meter-diameter mirror will allow to observe these scales. We used the 3 SPT-SZ and 6 \planck{}/HFI frequency maps (95, 150 and 220~GHz for SPT-SZ and 100, 143, 217, 353, 545 and 857~GHz for \planck{}/HFI) and studied a sample of 468 clusters from SPT-SZ~\cite{bocquet_cluster_2019} in the redshift range $0<z<1.7$.

\section{Analysis}

We ran the analysis pipeline on simulations of the 9 frequency maps and on real data. For real data, we cut $10 \times 10~\deg^2$ tangential maps from the All-sky maps for each frequency, and we directly build flat maps for the simulations.

\subsection{Simulations}

The simulations were kept simple, as a mean to detect possible biases in our analysis in a case with simple foregrounds. The first step was to create a $10 \times 10~\deg^2$ CMB map in $\Delta T/T$ units, from the \planck\ PR2 CMB TT spectrum. This map was first periodic, i.e. created in Fourier space and then transferred in real space. To make it more realistic, we also did the same with a $50 \times 50~\deg^2$ map and then cut a non periodic $10 \times 10~\deg^2$ map out of it. We then added the gravitational lensing effect from a cluster with a Navarro-Frenk-White (NFW) profile and its associated Sunyaev-Zel'dovich (SZ) effect for each frequency (for a total of 9 maps). The latter is how the CMB spectrum is modified because of the inverse Compton scattering of CMB photons by hot intracluster free electrons. It can have a thermal (tSZ) or kinetic origin (kSZ). Since kSZ has the same signature as the CMB, one can only separate the tSZ. The latter depends on the integrated electronic pressure, that we modelled with a generalised NFW profile, whereas the former was added only in systematic checks for cluster peculiar velocities of $v \sim 300$~km.s$^{-1}$. We then added the instrumental characteristics for each frequency, i.e. the beam, transfer function and noise.

\subsection{Pipeline}

For each cluster (real or simulated), we studied flat $10 \times 10~\deg^2$ maps at 9 different frequencies. Having the same patch of sky at different frequencies helps to clean it from foregrounds and noise. We used a constrained Internal Linear Combinations method~\cite{remazeilles_cmb_2011} (cILC) to get rid of the tSZ effect as well as mitigate the other contaminants. This method takes advantage of the fact we know the frequency signature of the tSZ effect as well as our instrumental characteristics. This way, we built the best lensed CMB map we could from our data. Then, we looked for a lensing signal at the cluster position using a quadratic estimator~\cite{hu_mass_2002} that extracts the lensing potential from the correlations between Fourier modes in the CMB. We then have information on a Fourier mode $\bm{K}=\bm{k'}-\bm{k}$ from the correlations between $\bm{k'}$ and $\bm{k}$. We finally compared the obtained lensing potential with the fiducial potential of a cluster with a NFW profile and the mass given in the SPT-SZ data set. To do so, we used a matched filter~\cite{melin_measuring_2015} that compares mode by mode the two profiles in Fourier space. The ratio of the profiles is equal to the ratio of the masses.

The real maps required a few additional steps: the sky map was projected on a tangential map using Healpy, the non periodic maps were apodised (the borders were smoothed out, also for the non periodic simulations) and the point sources were masked using a constrained Gaussian field~\cite{hoffman_constrained_1991} with the CMB properties and local continuity. Moreover, to subtract systematic biases due to the foregrounds, correlated noise or non periodicity, we ran the analysis pipeline on 10 empty field (\emph{off} map) for each field with a cluster (\emph{on} map). The final result is \emph{on} - $\langle \emph{off} \rangle_{10}$.

\section{Results}

\begin{figure}
    \centering
    \includegraphics[width=0.8\textwidth]{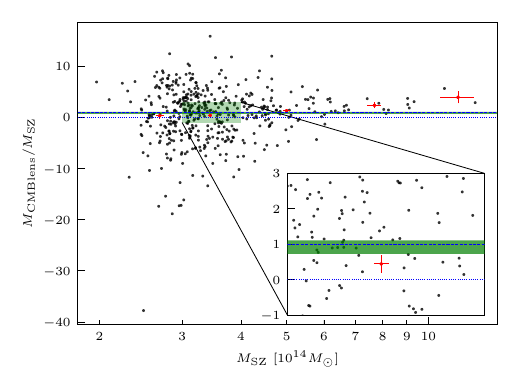}
    \caption{Recovered CMB lensing mass $M_{\rm CMBlens}$ (divided by the SZ mass $M_{\rm SZ}$) as a function of SZ mass for the \planck\ + SPT-SZ combined analysis. Black dots are individual measurements. The red dots are the weighted means of the five mass bins, spaced logarithmically. The dotted (resp. dashed) blue line is the zero (resp. unity) level.
      The green stripe is the weighted mean within the error
      $\langle M_{\rm CMBlens}/M_{\rm SZ} \rangle = 0.919 \pm 0.190$ (without correcting for the kSZ effect).}
    \label{fig:results}
\end{figure}

We obtained consistent results for \planck\ only, SPT-SZ only and the combination of both on periodic and non periodic simulations, and on real data~\cite{huchet_measurement_2024}. The combined result on real data is shown in Fig.\ref{fig:results}. There is an overall good agreement between our observed CMB lensing mass and the mass derived by SPT-SZ based on the tSZ signal. We observe a small increasing trend in the ratio between our mass measurement and the normalized SPT-SZ measurement with respect to the latter. This trend may be real or could be due to some remaining bias in our analysis. After correcting for the bias due to the kSZ estimated from the simulations ($\Delta \langle M_{\rm CMBlens}/M_{\rm SZ} \rangle = -0.06$), we find a final average mass ratio of $\langle \Mcmblens/\Msz \rangle = 0.98 \pm 0.19 \rm{~(stat.)} \pm 0.03 \rm{~(syst.)}$~\cite{huchet_measurement_2024}.

\begin{figure}
    \centering
    \includegraphics[width=0.85\textwidth]{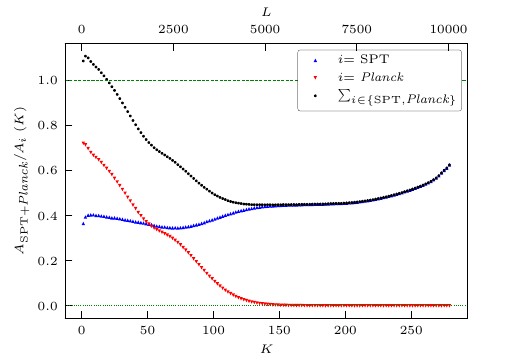}
    \caption{Comparison of the variance $\sigma^2(K)$ of the lensing
      profile reconstructed for SPT-SZ and \planck\, binned in
      $K=|\bm{K}|$ (or equivalently $L=|\bm{l'}-\bm{l}|$). The figure displays the ratios between the combined
      $\sigma_{\rm SPT+Planck}^2(K)$ and the single datasets $\sigma_i^2(K)$, and the sum of both ratios. See the main text for more information.
      }
    \label{fig:A_vs_K}
\end{figure}

We also observe that the statistical error bars gain a factor of $\sqrt{2}$ between one data set ($\sigma_{\planck} = 0.27$, $\sigma_{\rm SPT-SZ} = 0.29$) and the combination of both ($\sigma_{\rm combi} = 0.19$). It is not obvious that two data sets bring independent information when they observe the same sky. We show in Fig~\ref{fig:A_vs_K} the ratio $\sigma_{\rm combi}^2/\sigma_i^2$ with respect to the Fourier mode $K$, with $i$ representing either \planck\ or SPT-SZ. We also plotted the sum of both ratios for comparison with purely independent data sets. The figure shows that the \planck\ data set brings most of the information at large scales ($K<50.9$ or $L<1830$), while the SPT-SZ dataset provides information on the smaller scales ($K>145.0$ or $L>5208$). The sum is the product between the combined $\sigma_{\rm SPT+Planck}^2$ term and the inverse variance sum $1/\sigma_{\rm SPT}^2+1/\sigma_{\rm Planck}^2$. It is below unity for $K>18.9$ or $L>680$ demonstrating the significant gain of the combination with respect to individual measurements.

We therefore showed that information on CMB large scales is useful in the reconstruction of the lensing potential of a cluster using a quadratic estimator on temperature maps. In the future, \planck\ and/or the Small Aperture Telescopes (SATs) data might complement the Large Aperture Telescopes (LATs) on this matter.

\section*{References}

\bibliography{huchetalexandre.bib}

\end{document}